\documentclass[conference]{IEEEtran}
\IEEEoverridecommandlockouts
\usepackage{graphicx} % Required for inserting images
\usepackage{flushend}
\usepackage{balance}
\usepackage{url}
\usepackage{hyperref}
\usepackage{xurl}

\usepackage{float}
\usepackage{titlesec}

\begin{document}

\title{Farmer Connect: Improving Farmers' Access to Produce Markets
%\thanks{This work was carried out as part of our Bachelor of Computer Science  undergraduate research project in the Department of Computer Science, Mbarara University of Science and Technology, Uganda.}
}

\author{
\IEEEauthorblockN{
Amanya Micheal \IEEEauthorrefmark{1}, Kainamura Darius \IEEEauthorrefmark{1}, Namatovu Christine \IEEEauthorrefmark{1}, Kobugabe Lailah \IEEEauthorrefmark{1},\\ Buwule Solomon Fortune \IEEEauthorrefmark{1}, Adones Rukundo \IEEEauthorrefmark{1}
}
%, Theodora M. Twongyirwe \IEEEauthorrefmark{2}

\IEEEauthorblockA{
\IEEEauthorrefmark{1}
Department of Computer Science\\
Faculty of Computing and Informatics\\
Mbarara University of Science and Technology, Mbarara, Uganda\\
Email: \{2023bcs030,2023bcs056,2023bcs004,2023bcs002,2023bcs049\}@std.must.ac.ug, adones@must.ac.ug
}

%\IEEEauthorblockA{
%\IEEEauthorrefmark{2}
%Department of Mechanical and Industrial Engineering\\
%Faculty of Applied Science and Technology\\
%Mbarara University of Science and Technology, Mbarara, Uganda\\
%Email: tmtwongyirwe@must.ac.ug
%}
}

\maketitle

\begin{abstract}
Smallholder maize farmers in Uganda continue to face limited market access, weak bargaining power, low price transparency, and heavy reliance on intermediaries. These challenges are compounded by poor produce coordination, delayed payments, and weak visibility into cooperative transactions. This paper presents Farmer Connect, a cooperative-based digital platform designed to support produce management, marketplace coordination, and transparent earnings tracking among farmer groups. The system supports four user roles: administrators, supervisors, farmers, and customers. Its core functions include farmer group management, contribution recording and verification, marketplace listing, order processing, First In First Out based produce allocation, earnings visibility, mobile money payment support, and notification services. The platform was implemented using a mobile-first architecture with cloud-based backend services and an administrative web dashboard. Functional implementation showed that the system was able to support the major workflows required for group-based maize marketing and cooperative coordination, with approximately 85\% of identified user requirements implemented. The study shows that cooperative-centered digital platforms can provide a practical framework for improving transparency, coordination, and buyer access for smallholder farmers.
\end{abstract}

\begin{IEEEkeywords}
digital agriculture, smallholder farmers, maize marketing, cooperative systems, mobile marketplace, produce management, Uganda
\end{IEEEkeywords}

\section{Introduction}

Agriculture remains the backbone of Uganda’s economy, currently contributing approximately 24--26\% of the national Gross Domestic Product (GDP) and employing more than 65\% of the labor force \cite{npa2023, ubos2025}. The sector plays a central role in food security, household income generation, and rural livelihoods. Among Uganda’s major food crops, maize is the most widely cultivated staple and commercial crop, contributing significantly to national agricultural output and regional trade \cite{faostat2023}. Western Uganda, including districts such as Masindi, Kasese, Kamwenge, and Bundibugyo, is among the country’s leading maize-producing regions due to favourable agro-ecological conditions. Despite this potential, smallholder maize farmers continue to experience low productivity, limited market access, post-harvest losses, and unfair pricing practices \cite{suri2011, barrett2008}.

One of the major challenges affecting Uganda’s agricultural value chain is the weak linkage between farmers and markets. Many farmers market their produce individually through intermediaries commonly known as middle men who exploit information asymmetry and fragmented supply chains, resulting in low farm-gate prices \cite{worldbank2023, fao2022}. 

Smallholder Farmers  face a challenge of need for cash  after harvest  . Without timely access to cash, farmers are unable to purchase seeds, fertilizers, and labor needed to sustain productivity in the next planting season. This forces them to sell there produce to available market through intermediaries   hence ending up being cheated \cite{bergquist2018}

Agricultural cooperatives have historically served as institutional mechanisms for improving collective bargaining power, produce aggregation, and access to agricultural services. Currently Uganda has over 10,000 registered cooperatives operating across different agricultural sectors \cite{uca2022}. Studies show that cooperative participation can improve market access, strengthen extension services, and enhance farmers’ bargaining power \cite{fischer_qaim2012}. However, many cooperatives continue to face governance challenges such as weak accountability, poor produce tracking systems, financial opacity, and delayed earnings distribution \cite{wanyama2014, bernard_etal2008}.

The rapid growth of mobile technology and digital financial services presents opportunities for addressing these agricultural challenges. Mobile penetration in Uganda exceeds 67\%, while mobile money services have become widely accessible, with more than 43 million registered accounts \cite{gsma2022, bankofuganda2022}. Advances in cloud computing and cross-platform mobile application development have also reduced the cost of deploying digital systems in rural environments \cite{firebase2023}. Existing digital agricultural platforms such as Esoko, WeFarm, DigiFarm, and Twiga Foods have demonstrated the potential of mobile technologies in improving access to agricultural information and market coordination \cite{nakasone_etal2014, neven_etal2009, tsan_etal2019}. Nevertheless, most of these systems are designed primarily for individual farmers and do not adequately support cooperative-based produce management, transparent earnings distribution, and collective market operations.

To address these limitations, this study proposes Farmer Connect, a cooperative-centered digital platform for improving maize produce management and farmer-market connectivity in Uganda. Unlike existing digital marketplaces that support individual transactions, Farmer Connect enables farmers to aggregate produce and access markets collectively. The platform integrates mobile technology and mobile money services to support transparent interactions among farmers, supervisors, and buyers. Supervisors verify produce contributions, assign quality grades, and publish produce on a digital marketplace, while buyers can browse available stock, place orders, and access cooperative information. Farmers receive SMS notifications on produce verification, sales, and payments, enabling participation even without smartphones.

Farmer Connect uses automated FIFO stock allocation and transparent earnings calculation to improve accountability, reduce exploitation, and strengthen market access for smallholder farmers. The platform also supports financial inclusion by generating verifiable digital records of produce contributions, sales, and earnings, which can improve farmers’ access to loans, savings schemes, and government support programs such as the Parish Development Model (PDM). In addition, cooperative-based aggregation enables farmers to meet the quantity, quality, and traceability requirements of large buyers, including food processors, supermarket chains, and export agents. By acting as organized suppliers, cooperatives can negotiate larger contracts, access regional and international markets, and build long-term buyer relationships.

The platform further simplifies extension service delivery by allowing extension workers to engage directly with organized farmer groups. Farmer Connect aligns with Uganda’s Vision 2040 and the Sustainable Development Goals (SDGs), particularly Goal 2 (Zero Hunger), Goal 8 (Decent Work and Economic Growth), and Goal 9 (Industry, Innovation, and Infrastructure), by promoting inclusive agricultural commercialization and digital transformation in rural communities \cite{vision2040, sdgs}. This study therefore focuses on the design, development, and evaluation of Farmer Connect as a digital solution for cooperative-based maize produce management and market access in Uganda.

\section{Literature Review}

Smallholder farmers in Uganda continue to face persistent barriers to profitable market participation, including weak market linkages, limited bargaining power, poor access to timely market information, and dependence on intermediaries \cite{barrett2008, maaif2021, minot2013}. These constraints are particularly severe in fragmented produce markets such as maize, where farmers often sell individually, in small quantities, and under pressure to meet urgent household cash needs. In such conditions, produce is frequently sold at low prices to middlemen, reducing farmer incomes and weakening the benefits of agricultural production \cite{fafchamps_hill2005, omamo1998, sharma2020}.

Literature on collective action shows that farmer groups and cooperatives can improve bargaining power, reduce transaction costs, and strengthen access to markets, finance, and external support services \cite{fischer_qaim2012, bernard_etal2008, birchall_simmons2010}. However, the effectiveness of cooperatives depends heavily on governance quality, transparency, and accountability. In many sub-Saharan Africa cooperative settings, delayed payments, weak record keeping, and limited member visibility into transactions continue to undermine trust and participation \cite{wanyama2014, birchall2014}. These challenges are especially important in rural contexts, where farmers may rely on group structures not only for marketing, but also for social and financial support.

Digital agriculture research has shown that mobile platforms can improve market participation, advisory access, and information flow among smallholder farmers \cite{nakasone_etal2014, ayim2020, choruma2024}. Existing platforms such as Esoko, WeFarm, and DigiFarm demonstrate the value of mobile and SMS-based services in agricultural coordination \cite{tsan_etal2019}. However, most of these platforms focus on information access, advisory support, or individual market participation, rather than cooperative-centered produce aggregation, transparent earnings allocation, and group-based marketplace operations. As a result, important problems related to collective produce management, quality verification, and equitable revenue distribution remain insufficiently addressed.

Studies on agricultural supply chains further emphasize the importance of traceability, quality control, and inventory coordination in strengthening buyer confidence and commercialization outcomes \cite{handfield_nichols2002, christopher2005}. In Uganda’s maize sector, produce quality is commonly assessed using standards related to grain condition, moisture content, and impurities \cite{unbs2019}. Yet in practice, many smallholder farmers have limited visibility into how grading decisions are made, and this can reinforce information asymmetries between sellers and buyers \cite{jaffee_masakure2005}. Digital systems that capture produce records, grading outcomes, and transaction status can therefore improve transparency and reduce opportunities for manipulation.

Financial inclusion is also central to agricultural market participation. Evidence shows that mobile money can reduce transaction costs, speed up payments, and improve the security of financial transfers in low-resource settings \cite{mas_radcliffe2010, jack_suri2014, suri_jack2016}. For rural farming households, digital payments may also strengthen control over earnings and reduce leakages associated with informal cash handling. 
%This is particularly relevant where women participate actively in production but may have limited control over proceeds from produce sales.
This is particularly relevant in contexts where women participate actively in agricultural production but have limited control over proceeds from produce sales.
In addition, organized digital records may support future access to credit, extension services, and targeted government interventions by making farmer activity and produce flows more visible \cite{karlan_etal2014, ferris2014}.

At the same time, adoption of digital platforms in rural agriculture depends on affordability, usability, and compatibility with existing institutional arrangements \cite{mumford2006, aker_mbiti2010, gsma2022}. Offline-capable mobile systems, role-based access control, and simple user interfaces are especially important in contexts characterized by intermittent internet connectivity and varied digital literacy levels \cite{firebase2023, sandhu_etal1996, ferraiolo_etal2001}. These considerations suggest that effective agricultural systems should not only digitize transactions, but should also reinforce the social and operational realities of farmer groups.

Overall, the literature shows that although digital agriculture platforms have improved information access, communication, and selected market functions, significant gaps remain in cooperative-based produce management, transparent earnings distribution, and group-centered market coordination for maize farmers in Uganda. Farmer Connect was therefore conceived to address this gap by integrating farmer grouping, produce recording, quality verification, marketplace coordination, earnings transparency, and digital payment support within a single platform tailored to the needs of smallholder maize cooperatives.

\section{Platform Design}

This section presents the system development process of Farmer Connect, focusing on the identified requirements, system architecture, workflows, and operational processes implemented to support cooperative-based maize produce management and market access.

\subsection{Requirements Identified}

The requirements for Farmer Connect were derived from consultations with farmers, cooperative supervisors, buyers, and administrators. The platform was designed to support four major categories of users: administrators, supervisors, farmers, and customers. Each role was assigned specific responsibilities within the cooperative-based agricultural marketplace.

\subsubsection{User Requirements}

\emph{Administrator Requirements:} The administrator role supports overall platform governance and coordination of cooperative activities within Farmer Connect. The platform enables administrators to register and manage farmer cooperative groups, assign supervisors to specific groups, and maintain user records across the platform. Through a centralized web-based dashboard, administrators can monitor produce submissions, track customer orders, review transaction records, and oversee payment-related activities. The administrator also generates analytical reports to support cooperative performance evaluation, produce movement tracking, and revenue monitoring. These functions are essential for ensuring accountability, transparency, and effective coordination across the digital agricultural marketplace.

\emph{Supervisor Requirements:} Supervisors serve as the operational coordinators of the cooperative and are responsible for overseeing produce intake and marketplace activities. The platform enables supervisors to register farmers under their assigned cooperative groups, record produce contributions, upload photographic evidence, assign quality grades, and verify or reject submitted produce. In addition, supervisors can publish approved produce to the marketplace, manage customer orders, and monitor group-level statistics and membership records. These functionalities support accurate produce management, operational accountability, and effective coordination within the cooperative.

\emph{Farmer Requirements:} Farmers use the platform primarily to track their produce contributions, earnings, and participation within the cooperative. The platform allows farmers to view their contribution history, verification status, earnings calculations, proportional share within the cooperative, pending balances, and available balances. It also provides notifications related to contribution approval, produce sales, and payment updates. These features enhance transparency, improve farmers’ awareness of cooperative operations, and strengthen trust in the produce management and payment process.

\emph{Customer Requirements:} The customer role enables buyers to access produce offered by registered farmer groups through the Farmer Connect marketplace. The platform allows customers to register using their phone numbers, browse available produce listings organized by cooperative groups, and review key product details such as quantity, pricing, and images. Customers can place orders, monitor order progress, receive transaction updates, and complete payments through supported mobile money services. These functions improve market access, streamline buyer-seller interaction, and support a more transparent produce trading process.

\subsubsection{Functional Requirements}

 The functional requirements of Farmer Connect were organized into major components that support secure access, cooperative coordination, produce management, and transparent benefit sharing within farmer groups.

%\begin{enumerate}
    \emph{Authentication and User Management:} 
    The platform provides role-based authentication for administrators, supervisors, farmers, and customers. Administrators, supervisors, and farmers authenticate using email and password credentials, while customers register and access the platform using phone numbers. Role-based access control is implemented to restrict each user to functions relevant to their responsibilities, while persistent session management supports continued access to authorized services. This ensures secure and structured participation across the platform.

    \emph{Group and Cooperative Management:}
    The platform supports the creation and management of farmer groups and cooperative structures. It enables the registration of groups, assignment of supervisors, and maintenance of group membership records. The platform also tracks cooperative-level indicators such as aggregated produce volume, membership, and revenue performance. This functionality is important for organizing smallholder farmers into collective units that can coordinate production, strengthen market participation, and improve bargaining power.

    \emph{Contribution Recording and Verification:}
    The platform enables supervisors to record produce delivered by farmers, including quantity, images, and quality grade classifications such as A, B, and C. Recorded contributions are then reviewed and either verified or rejected before being made available for market access. Only verified produce is published to the marketplace, thereby supporting product quality assurance, strengthening buyer confidence, and improving the integrity of cooperative-based trading.

    \emph{Earnings and Share Calculation:}
    The platform computes farmer earnings and proportional share contributions within their respective groups. Earnings are classified into pending and available balances depending on the sale status of contributed produce. Share calculations are used to maintain fair attribution of value among participating farmers based on their recorded contributions. This functionality promotes transparency in benefit distribution and helps farmers track the financial outcomes of their participation in collective marketing.

    \emph{Marketplace and Order Processing:}
    The platform provides a cooperative-based digital marketplace through which verified produce is made available to customers. It enables customers to browse produce listings, place orders, and track order progress. To promote fairness in produce allocation and minimize storage-related losses, the platform processes orders using a First-In-First-Out (FIFO) algorithm that prioritizes older verified contributions.

    \emph{Payment and Settlement:}
    The platform integrates mobile money payment services to support secure and timely financial transactions within the cooperative. In addition to settling farmer earnings after produce has been sold, the platform supports advance payments made at the discretion of the cooperative. This allows cooperatives to pay farmers before the final sale of their produce, particularly in situations where produce is held temporarily in anticipation of more favorable market prices. The platform records such payments as advances against expected earnings and, once the produce is sold, reconciles the advance with the actual sale proceeds, deducts the amount already paid, and computes the remaining balance payable to the farmer. All settlement records and payment histories are maintained to promote transparency, accountability, and traceability within the cooperative.

    \emph{Notifications and Reporting:}
    Farmer Connect provides real-time notifications to keep users informed about key activities such as contribution verification, order updates, produce sales, and payment transactions. In addition, the system offers administrative reporting tools that support performance monitoring, trend analysis, and data export for informed decision-making and effective cooperative management.
%\end{enumerate}

\subsection{System Architecture}

Farmer Connect was designed as a mobile-first, cloud-supported platform for cooperative-based maize produce management and market access. The architecture is organized into three layers: the presentation layer, application layer, and data layer.

The presentation layer comprises a mobile application for farmers, supervisors, and customers, and a web-based dashboard for administrators. This layer enables user interaction with the system, including produce submission, marketplace access, order placement, and administrative oversight.

The application layer implements the core business processes of the platform. These include user authentication, contribution verification, produce listing, order processing, earnings computation, notification management, and payment-related operations.

The data layer manages the storage and synchronization of system records, including user profiles, cooperative groups, produce contributions, orders, payments, and transaction histories. Supporting cloud services are used for authentication, image storage, and secure financial transactions..

\subsection{System Flow}

Farmer Connect implements operational workflows that coordinate interactions among farmers, supervisors, customers, and administrators within the cooperative-based marketplace.

\begin{figure*}[!t]
\centering
\includegraphics[
trim=0cm 0.36cm 0cm 0.6cm,
clip,
width=\textwidth,
height=0.9\textheight,
keepaspectratio
]{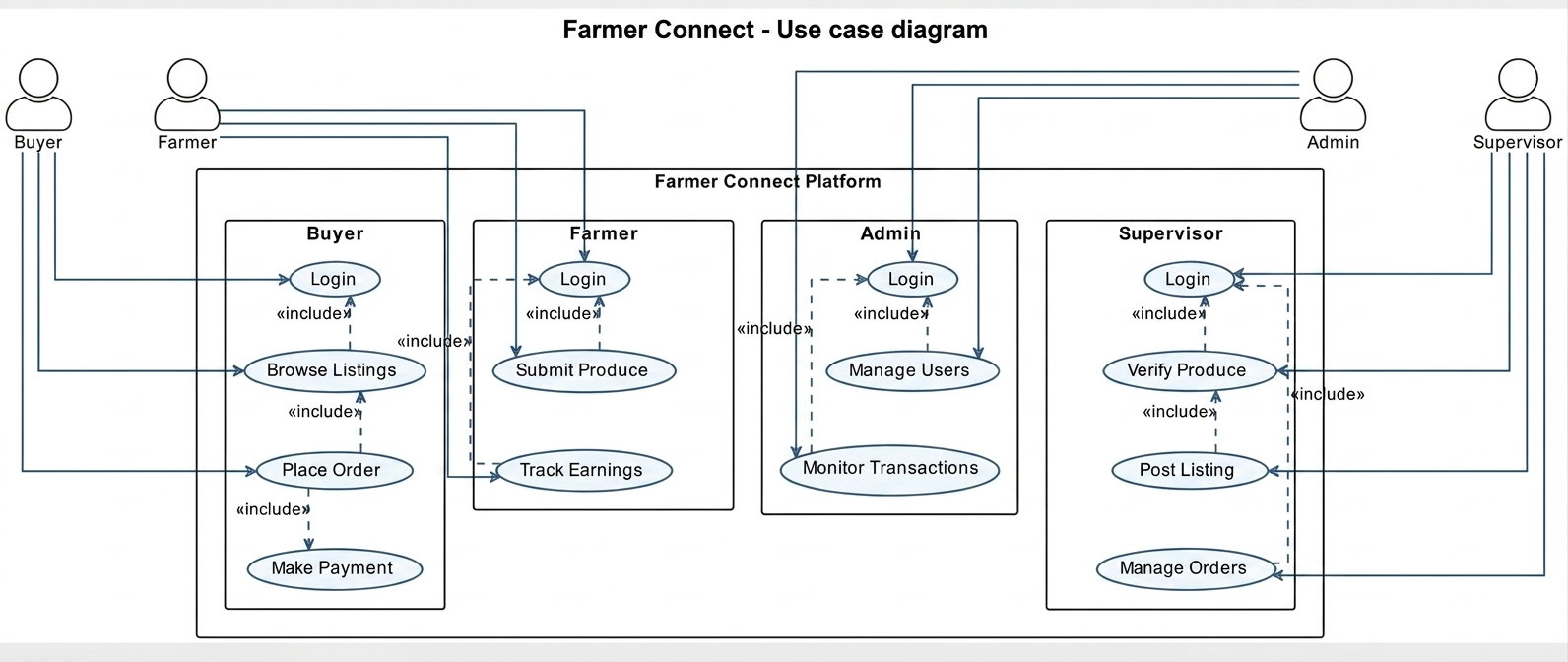}
\caption{Use Case Diagram}
\label{fig:usecase-diagram}
\end{figure*}

\subsubsection{Farmer Onboarding Flow}

The onboarding process begins with the creation of a farmer group and the assignment of a supervisor. The supervisor then registers farmers under the assigned group and provides them with access credentials. Once enrolled, farmers can use the mobile application to view their contribution records, receive updates, and track earnings associated with group-based produce sales.

\subsubsection{Contribution Recording and Verification Flow}

When farmers deliver produce to the group collection point, the supervisor records the contribution using the mobile application. The recorded details include quantity, quality grade, price per unit, and supporting images where applicable. Each contribution is initially marked as pending and is subjected to verification before market listing. Verified contributions are published to the marketplace, while rejected contributions are excluded and recorded accordingly to maintain quality control and marketplace reliability.

\subsubsection{Marketplace Ordering and First-In-First-Out (FIFO) Allocation Flow}

Customers access the marketplace to browse verified produce listings organized by cooperative group. Once a customer places an order, the system applies a FIFO allocation procedure that prioritizes the oldest verified produce contributions. Where a single contribution is insufficient to satisfy the requested quantity, the system allocates produce across multiple verified contributions in chronological order. This workflow promotes fairness among farmers, improves stock rotation, and reduces the risk of spoilage or prolonged storage. After allocation, the order is forwarded for confirmation and fulfillment by the responsible supervisor.

\subsubsection{Payment and Settlement Flow}

Following order confirmation and fulfillment, the platform supports payment processing through mobile money services. The platform is designed to accommodate two settlement approaches. In the first approach, farmers receive payment after their produce has been sold. In the second approach, the cooperative may choose to make advance payments to farmers before the produce is sold, particularly when produce is being held in anticipation of better market prices. In such cases, the system records the payment as an advance against the farmer’s expected earnings. Once the produce is eventually sold, the platform reconciles the advance with the actual sale proceeds, recovers the amount previously paid by the cooperative, and computes any remaining balance due to the farmer. Notifications are then issued to farmers to confirm payment updates and settlement status.

\subsubsection{Earnings Calculation and Share Distribution}

The platform continuously calculates farmer earnings based on the value of verified produce contributions and the applicable pricing model. It also computes each farmer’s proportional share within the cooperative according to the quantity and value of produce contributed relative to the total verified stock. Earnings are categorized into pending and available balances depending on the settlement stage, while any advance payments made by the cooperative are recorded and considered during final reconciliation. These real-time calculations improve transparency, enable farmers to monitor their financial position, and support fair distribution of proceeds within the cooperative.

\subsection{Use Case Design}

The Farmer Connect use case model defines four primary actors who interact with the system to support cooperative-based produce management and market access as illustrated in Figure \ref{fig:usecase-diagram}.

%\begin{itemize}
    \emph{Administrator:} oversees platform operations, manages cooperative groups and supervisors, and monitors system performance through administrative reports and analytics.
    
    \emph{Supervisor:} manages farmer group activities by registering farmers, recording and verifying produce contributions, publishing verified produce, and coordinating order fulfillment.

    \emph{Farmer:} interacts with the platform to monitor produce contributions, verification outcomes, earnings, payment balances, and share participation within the cooperative. The farmer also receives notifications related to produce sales, payment updates, and other relevant transactions.

    \emph{Customer:} uses the platform to browse verified produce listings, access product and cooperative details, place orders, monitor order progress, and complete transactions using the supported mobile money payment channels.
%\end{itemize}

The use case interactions collectively define the operational boundaries of the Farmer Connect platform and ensure proper role separation within the cooperative ecosystem.

\section{Results and Discussion}
This study focused on the design and implementation of Farmer Connect as a cooperative-based digital platform for improving maize produce management, buyer access, and earnings transparency among smallholder farmers in Uganda. Rather than treating digital agriculture as a purely technical intervention, the platform was designed around challenges identified in the literature, including fragmented market participation, weak produce coordination, limited transparency in cooperative transactions, delayed payments, and dependence on intermediaries.

One of the central design decisions in Farmer Connect was the organization of farmers into supervised groups rather than treating them as isolated individual sellers. This approach responds to findings in the literature that smallholder farmers often face weak bargaining power, low market visibility, and high transaction costs when they sell independently. By structuring the system around farmer groups, the platform supports produce aggregation, coordinated marketing, and stronger buyer-facing visibility. 

\begin{figure}[htbp]
\centering
\includegraphics[width=1\linewidth]{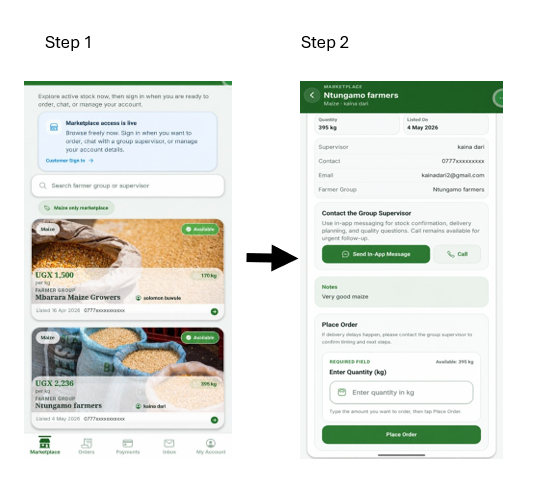}
\caption{Marketplace interface showing produce listings from farmer groups.}
\label{fig:marketplace}
\end{figure}

As illustrated in Figure \ref{fig:marketplace}, the marketplace module allows customers to browse produce offered by organized groups, review available listings, and place orders through a structured workflow. This creates a direct digital interaction between buyers and farmer groups, which may reduce reliance on informal middlemen and improve access to broader markets.

\begin{figure}[htbp]
\centering
\includegraphics[width=0.7\linewidth]{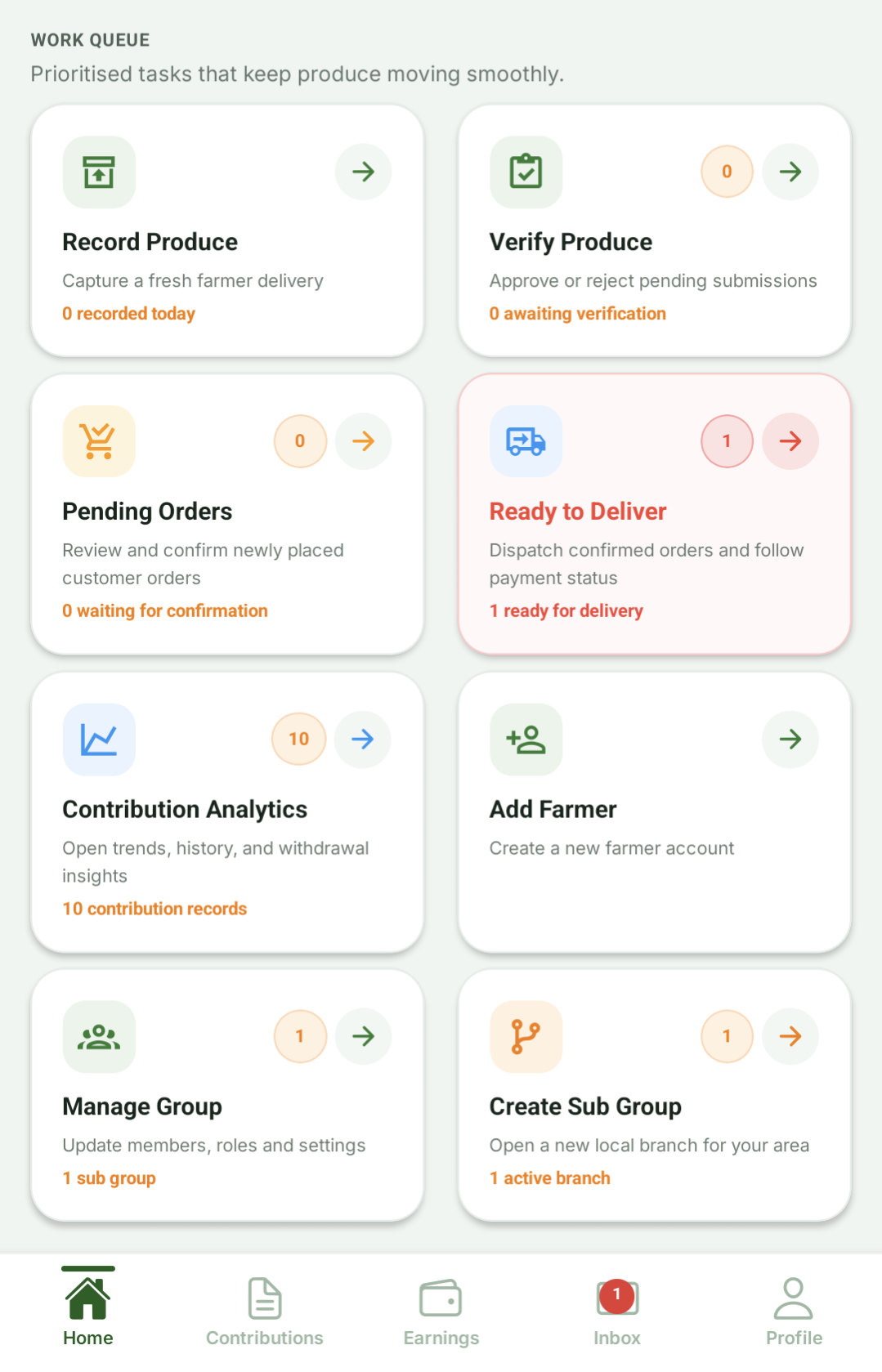}
\caption{Supervisor interface .}
\label{fig:supervisor}
\end{figure}

The supervisor module was included to address practical realities within rural cooperative operations. In many farmer groups, produce is received at collection points and must be inspected, recorded, and validated before it can be offered for sale. Farmer Connect therefore provides supervisors with tools for contribution recording, quality grading, image capture, and contribution verification as illustrated in Figure \ref{fig:supervisor}. This workflow supports traceability and quality assurance, while ensuring that only verified produce is listed on the marketplace. Such functionality is important in contexts where poor record keeping and informal grading practices can create mistrust among farmers and uncertainty for buyers.

\begin{figure}[htbp]
\centering
\includegraphics[width=0.9\linewidth]{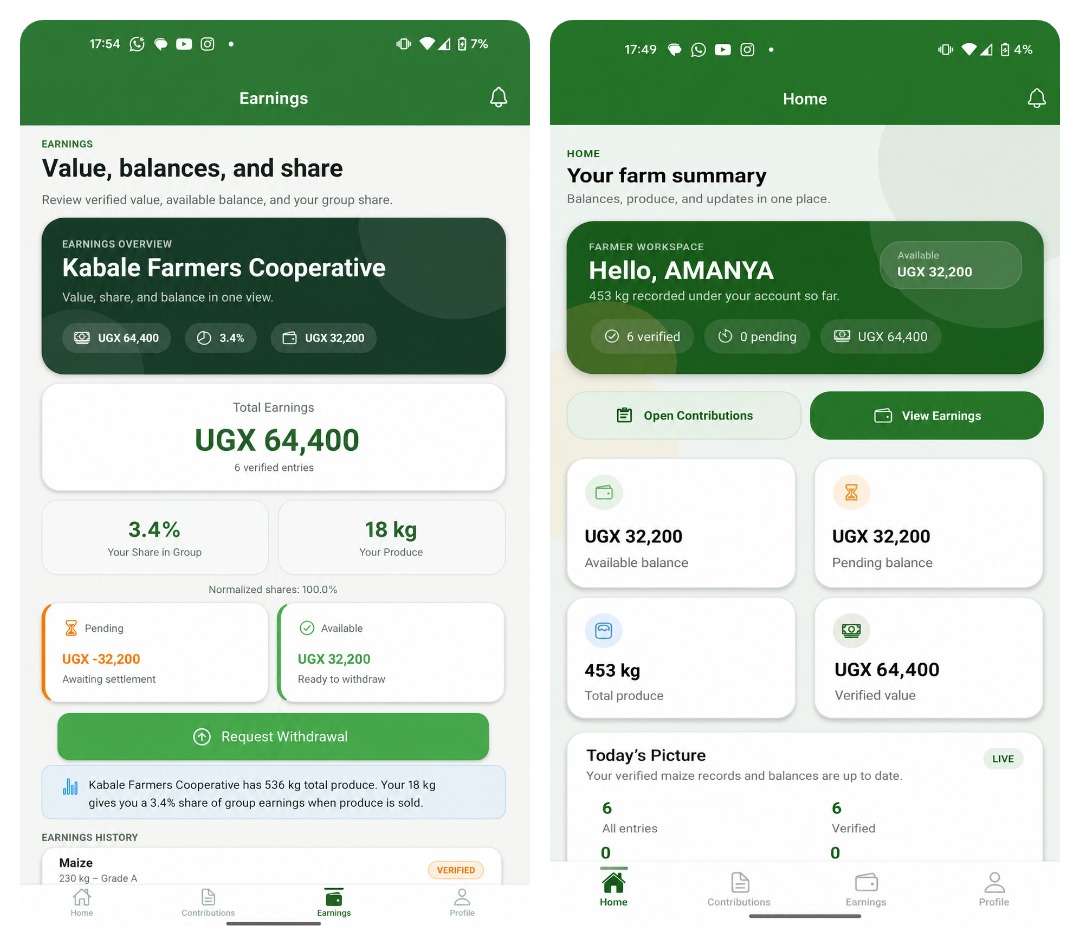}
\caption{Farmer interfaces showing their statistics.}
\label{fig:farmerscreens}
\end{figure}

The platform was also designed to improve transparency within cooperative operations. Literature reviewed in this study highlighted the persistent problem of delayed payments, opaque calculations, and limited farmer visibility into group transactions. In response, as illustrated in Figure \ref{fig:farmerscreens}, Farmer Connect enables farmers to monitor contribution history, verification status, and earnings records through the mobile application. This is intended to reduce disputes associated with manual record management and to strengthen accountability between members and cooperative leaders. The inclusion of proportional share tracking and earnings visibility was particularly important because transparent benefit allocation is central to sustaining trust in collective marketing systems.

Another important design consideration was the financial pressure that often forces farmers to sell produce immediately after harvest at unfavourable prices. By supporting group-based produce management and delayed marketplace sale, the platform creates a structure through which produce can be held, verified, and marketed more strategically. While this study did not measure long-term income effects, the design creates a practical basis for reducing distress sales driven by urgent cash needs. In this sense, the platform is not only a transaction tool, but also a coordination mechanism that may support stronger collective decision-making around when and how produce is sold.

Mobile money integration was included to address the challenge of delayed and insecure payment processes. The literature shows that digital financial services can improve transaction speed, reduce cash handling risks, and support financial inclusion in rural economies. In Farmer Connect, payment support was integrated into the transaction lifecycle so that order completion and payment records remain visible within the same platform environment. This improves transaction traceability and may strengthen confidence among both farmers and buyers. In addition, digital payment records may provide a foundation for future integration with savings, credit, or other rural financial support services.

The platform also considered the technological constraints of rural communities. Because not all group members are expected to own smartphones or maintain reliable internet access. We incorporate SMS notification-based communication to extend visibility beyond the primary app user. This design choice is important because it broadens inclusion within the group and helps ensure that farmers remain informed about contribution verification, produce sales, and payment updates even where device ownership is uneven. The use of mobile-first workflows, cloud synchronization, and lightweight user interactions further reflects the need to balance functionality with accessibility in low-resource settings.

At the administrative level, Farmer Connect is designed to support platform governance, cooperative oversight, and transaction monitoring. The administrator dashboard enables management of groups, supervisors, records, and high-level operational activity. This functionality is relevant not only for platform control, but also for strengthening accountability and organized record keeping across the cooperative network. More broadly, such structured records could support extension coordination, program monitoring, and evidence-based intervention planning where organized farmer groups are involved.

Despite these strengths, some practical limitations remain. Effective use of the platform depends on reliable participation by supervisors, adequate digital literacy, and sufficient connectivity for synchronization and notifications. In addition, although the system improves transparency, it does not completely eliminate institutional risks such as weak leadership or misuse of authority within groups. Strong cooperative governance and user training therefore remain necessary for the platform to function effectively in practice.

Overall, the discussion suggests that Farmer Connect was designed in a way that aligns closely with the problems identified in the literature. Its major functions, including group-based organization, produce verification, direct marketplace access, earnings visibility, mobile money support, and administrative oversight, were intentionally selected to respond to known weaknesses in maize marketing and cooperative operations among smallholder farmers. The platform therefore offers a practical digital model for improving coordination, transparency, and market participation within Uganda's cooperative agricultural context. 

\section{Conclusion}
The development of Farmer Connect focused on the design and implementation of a cooperative-based digital platform for maize produce management and market access. The platform integrates role-based access control, cloud-supported service delivery, produce verification, order processing, earnings tracking, mobile money support, and SMS notifications to enhance transparency, accountability, and operational efficiency within agricultural cooperatives. Through collective digital marketing and produce aggregation, the platform enables smallholder farmers to approach larger markets as organized groups, thereby improving bargaining power, reducing transportation costs, and strengthening access to reliable buyers.

The platform further improves cooperative governance through transparent produce tracking, quality grading, FIFO stock allocation, and digital earnings management, reducing disputes and opportunities for exploitation. By generating verifiable digital records of produce contributions, sales, and earnings, Farmer Connect also promotes financial inclusion by supporting farmers’ access to agricultural loans, savings schemes, government support programs, and other financial services. Overall, the system demonstrates the potential of digital technologies in strengthening agricultural commercialization, market transparency, and sustainable socio-economic development among smallholder maize farmers in Uganda.

\bibliographystyle{ieeetr}
\bibliography{references}

\begin{thebibliography}{10}

\bibitem{npa2023}
{National Planning Authority}, ``Policy paper on transforming smallholder farming to modern agriculture in uganda.'' \url{https://www.npa.go.ug/wp-content/uploads/2023/03/NPA-PEC-PAPER-TRANSFORMING-SMALLHOLDER-FARMING-TO-MODERN-AGRICULTURE-IN-UGANDA.pdf}, 2023.
\newblock Accessed: 28 October 2025.

\bibitem{ubos2025}
U.~B. of~Statistics, ``Preliminary annual gross domestic product 2024/25,'' 2025.
\newblock \url{https://www.ubos.org/preliminary-annual-gross-domestic-product-2024-25/}.

\bibitem{faostat2023}
{FAOSTAT}, ``Crops and livestock products database: {Uganda} maize production.'' Food and Agriculture Organisation. Available at: \url{https://www.fao.org/faostat}, 2023.
\newblock Accessed: 15 February 2025.

\bibitem{suri2011}
T.~Suri, ``Selection and comparative advantage in technology adoption,'' {\em Econometrica}, vol.~79, no.~1, pp.~159--209, 2011.

\bibitem{barrett2008}
C.~B. Barrett, ``Smallholder market participation: {C}oncepts and evidence from eastern and southern {Africa},'' {\em Food Policy}, vol.~33, no.~4, pp.~299--317, 2008.

\bibitem{worldbank2023}
W.~Bank, ``Uganda agricultural market systems report 2023,'' 2023.

\bibitem{fao2022}
Food and A.~O. of~the United~Nations, ``Fao country profile: Uganda,'' 2022.

\bibitem{bergquist2018}
L.~F. Bergquist and C.~McIntosh, ``Building market linkages for smallholder farmers through a digital marketplace in uganda,'' tech. rep., Innovations for Poverty Action, 2018.

\bibitem{uca2022}
{UCA}, ``Uganda cooperative alliance annual report 2021/22,'' tech. rep., Uganda Cooperative Alliance, Kampala, 2022.

\bibitem{fischer_qaim2012}
E.~Fischer and M.~Qaim, ``Linking smallholders to markets: {D}eterminants and impacts of farmer collective action in {Kenya},'' {\em World Development}, vol.~40, no.~6, pp.~1255--1268, 2012.

\bibitem{wanyama2014}
F.~O. Wanyama, ``Cooperatives and the sustainable development goals: {A} contribution to the post-2015 development debate,'' tech. rep., International Labour Organisation, Geneva, 2014.

\bibitem{bernard_etal2008}
T.~Bernard, M.-H. Collion, A.~De~Janvry, P.~Rondot, and E.~Sadoulet, ``Do village organizations make a difference in {African} rural development? {A} study for {Senegal} and {Burkina Faso},'' {\em World Development}, vol.~36, no.~11, pp.~2188--2204, 2008.

\bibitem{gsma2022}
{GSMA Intelligence}, ``The mobile economy: {S}ub-{S}aharan {Africa} 2022,'' tech. rep., GSMA, London, 2022.

\bibitem{bankofuganda2022}
{Bank of Uganda}, ``Annual report 2021/22: {F}inancial inclusion and mobile money statistics,'' tech. rep., Bank of Uganda, Kampala, 2022.

\bibitem{firebase2023}
{Firebase}, ``Cloud {Firestore} documentation.'' Google {LLC}. Available at: \url{https://firebase.google.com/docs/firestore}, 2023.
\newblock Accessed: 10 March 2025.

\bibitem{nakasone_etal2014}
E.~Nakasone, M.~Torero, and B.~Minten, ``The power of information: {T}he {ICT} revolution in agricultural development,'' {\em Annual Review of Resource Economics}, vol.~6, no.~1, pp.~533--550, 2014.

\bibitem{neven_etal2009}
D.~Neven, M.~M. Odera, T.~Reardon, and H.~Wang, ``{Kenyan} supermarkets, emerging middle-class horticultural farmers, and employment impacts on the rural poor,'' {\em World Development}, vol.~37, no.~11, pp.~1802--1811, 2009.

\bibitem{tsan_etal2019}
M.~Tsan, S.~Totapally, M.~Hailu, and N.~J. Sitko, ``The digitalisation of {African} agriculture report 2018--2019,'' tech. rep., CTA and Dalberg Advisors, Wageningen, 2019.

\bibitem{vision2040}
G.~of~Uganda, ``Uganda vision 2040,'' 2013.
\newblock Revised 2020.

\bibitem{sdgs}
``Sustainable development goals,'' n.d.
\newblock Accessed October 16, 2025.

\bibitem{maaif2021}
{MAAIF}, ``Agricultural sector strategic plan 2020/21--2024/25,'' tech. rep., Ministry of Agriculture, Animal Industry and Fisheries, Republic of Uganda, Kampala, 2021.

\bibitem{minot2013}
N.~Minot and R.~V. Hill, ``Developing and connecting markets for poor farmers,'' {\em IFPRI Policy Brief}, 2013.

\bibitem{fafchamps_hill2005}
M.~Fafchamps and R.~V. Hill, ``Selling at the farmgate or traveling to market,'' {\em American Journal of Agricultural Economics}, vol.~87, no.~3, pp.~717--734, 2005.

\bibitem{omamo1998}
S.~W. Omamo, ``Farm-to-market transaction costs and specialisation in small-scale agriculture,'' {\em Journal of Development Studies}, vol.~35, no.~2, pp.~152--163, 1998.

\bibitem{sharma2020}
A.~Sharma {\em et~al.}, ``Problems faced by farmers in traditional markets,'' {\em Journal of Agricultural Economics and Development}, vol.~9, no.~4, pp.~56--69, 2020.

\bibitem{birchall_simmons2010}
J.~Birchall and R.~Simmons, ``The role of co-operatives in poverty reduction: {N}etwork perspectives,'' {\em Journal of Socio-Economics}, vol.~39, no.~4, pp.~247--252, 2010.

\bibitem{birchall2014}
J.~Birchall, {\em The Governance of Large Co-operative Businesses}.
\newblock Manchester: Co-operatives {UK}, 2014.

\bibitem{ayim2020}
C.~Ayim, A.~Kassahun, B.~Tekinerdogan, and C.~Addison, ``Adoption of ict innovations in the agriculture sector in africa: A systematic literature review,'' {\em Agriculture \& Food Security}, vol.~9, no.~1, p.~34, 2020.

\bibitem{choruma2024}
D.~J. Choruma and T.~L. Dirwai, ``Digitalisation in agriculture: A scoping review of technologies in practice, challenges, and opportunities for smallholder farmers in sub-saharan africa,'' {\em Journal of Agriculture and Food Research}, vol.~18, pp.~8--101286, July 2024.

\bibitem{handfield_nichols2002}
R.~B. Handfield and E.~L. Nichols, {\em Supply Chain Redesign: {T}ransforming Supply Chains into Integrated Value Systems}.
\newblock Upper Saddle River: Financial Times Prentice Hall, 2002.

\bibitem{christopher2005}
M.~Christopher, {\em Logistics and Supply Chain Management}.
\newblock Harlow: Pearson Education, 3rd~ed., 2005.

\bibitem{unbs2019}
{UNBS}, ``Uganda standard: {M}aize grain --- specification ({US} 46:2019),'' tech. rep., Uganda National Bureau of Standards, Kampala, 2019.

\bibitem{jaffee_masakure2005}
S.~Jaffee and O.~Masakure, ``Strategic use of private standards to enhance international competitiveness,'' {\em Food Policy}, vol.~30, no.~3, pp.~316--333, 2005.

\bibitem{mas_radcliffe2010}
I.~Mas and D.~Radcliffe, ``Mobile payments go viral: {M-Pesa} in {Kenya},'' {\em Capco Institute Journal of Financial Transformation}, vol.~32, pp.~169--182, 2010.

\bibitem{jack_suri2014}
W.~Jack and T.~Suri, ``Risk sharing and transactions costs: {E}vidence from {Kenya's} mobile money revolution,'' {\em American Economic Review}, vol.~104, no.~1, pp.~183--223, 2014.

\bibitem{suri_jack2016}
T.~Suri and W.~Jack, ``The long-run poverty and gender impacts of mobile money,'' {\em Science}, vol.~354, no.~6317, pp.~1288--1292, 2016.

\bibitem{karlan_etal2014}
D.~Karlan, R.~Osei, I.~Osei-Akoto, and C.~Udry, ``Agricultural decisions after relaxing credit and risk constraints,'' {\em Quarterly Journal of Economics}, vol.~129, no.~2, pp.~597--652, 2014.

\bibitem{ferris2014}
S.~Ferris, E.~Kaganzi, and et~al., ``Linking smallholder farmers to markets and the implications for extension and advisory services,'' {\em Agricultural Systems}, vol.~123, pp.~53--62, 2014.
\newblock Explores market linkages, extension services, and impacts on smallholder farmer income.

\bibitem{mumford2006}
E.~Mumford, ``The story of socio-technical design: {R}eflections on its successes, failures and potential,'' {\em Information Systems Journal}, vol.~16, no.~4, pp.~317--342, 2006.

\bibitem{aker_mbiti2010}
J.~C. Aker and I.~M. Mbiti, ``Mobile phones and economic development in africa,'' {\em Journal of Economic Perspectives}, vol.~24, no.~3, pp.~207--232, 2010.

\bibitem{sandhu_etal1996}
R.~S. Sandhu, E.~J. Coyne, H.~L. Feinstein, and C.~E. Youman, ``Role-based access control models,'' {\em IEEE Computer}, vol.~29, no.~2, pp.~38--47, 1996.

\bibitem{ferraiolo_etal2001}
D.~F. Ferraiolo, R.~Sandhu, S.~Gavrila, D.~R. Kuhn, and R.~Chandramouli, ``Proposed {NIST} standard for role-based access control,'' {\em ACM Transactions on Information and System Security}, vol.~4, no.~3, pp.~224--274, 2001.

\end{thebibliography}

\end{document}